**Autonomous waves and global motion modes in living active solids**

Haoran Xu [1], Yulu Huang [2], Rui Zhang [2], Yilin Wu [1*]

[1] *Department of Physics and Shenzhen Research Institute, The Chinese University of Hong Kong, Shatin, NT, Hong Kong, P.R. China.*
[2] *Department of Physics, The Hong Kong University of Science & Technology, Clear Water Bay, Kowloon, Hong Kong, P.R. China.*

*To whom correspondence should be addressed.  Mailing address: Room 306, Science Centre, Department of Physics, The Chinese University of Hong Kong, Shatin, NT, Hong Kong, P.R. China.  Tel: (852) 39436354.  Fax: (852) 26035204.  Email: ylwu@cuhk.edu.hk

**Abstract**:
Elastic active matter or active solid consists of self-propelled units embedded in an elastic matrix.  Active solid resists deformation; the shape-preserving property and the intrinsic non-equilibrium nature make active solids a superior component for self-driven devices.  Nonetheless, the mechanical properties and emergent behavior of active solids are poorly understood.  Using a biofilm-based bacterial active solid, here we discovered self-sustained elastic waves with unique wave properties not seen in passive solids, such as power-law scaling of wave speed with activity.  Under isotropic confinement, the active solid develops two topologically distinct global motion modes that can be selectively excited, with a surprising step-like frequency jump at mode transition.  Our findings reveal novel spatiotemporal order in elastic active matter and may guide the development of solid-state adaptive or living materials.

**Text:**

Active matter consists of living or synthetic units that can convert local free energy input to mechanical work [1]. As an intrinsically non-equilibrium state of matter, active matter displays rich emergent behavior such as spontaneous collective motion [2-4] and self-organization [5-12]; it holds great potential for understanding the physics of living matter [1] and for the design of novel autonomous or self-driven materials with life-like properties [13,14]. Understanding the mechanical functionality of active matter as a continuum (i.e. the ability to generate forces or extract work at length scales much larger than the individual unit) is an essential step to bridge the gap between practical applications and a myriad of fundamental science knowledge in the field learned over the past two decades [15]. However, such understanding has been limited, in part due to the poor scalability of most experimental systems. Here we use millimeter-sized bacterial biofilms [16,17] to explore the emergent mechanical behavior of elastic active matter or *active solids* [18-22]. We discovered that mass elements in bacterial active solids are self-driven into local oscillatory motion. Under two-dimensional isotropic confinement, the local oscillation self-organizes into a pair of topologically distinct global motion modes. The mode selection is tunable by varying the activity of the active units; surprisingly, the two modes transit between each other with a sharp, step-like frequency jump at certain activity threshold. Under anisotropic confinement with a major axis of symmetry, the local oscillation of mass elements is organized in space as self-sustained elastic standing waves. These results are observed in both experiments and numerical modeling with remarkable agreement. Furthermore, our model predicts that the phase of local oscillations is arranged as traveling waves in unconfined space, with the wave speed scaling as the ~1/2 power of activity and the wavelength independent of frequency. This property of active matter elastic waves is in stark contrast to the counterpart in passive mechanical waves where the wave speed does not depend on the driving amplitude of external stimuli. Our findings reveal novel mechanical properties of active matter and pave the way for investigating the non-equilibrium physics of elastic active matter in *continuum*. The findings may guide the development of solid-state adaptive or living materials, such as autonomous actuators for soft robotics [23,24], programmable tissues [25,26], and synthetic microbial consortia [27,28] with mechanical functionalities.

Elastic active matter consists of force-generating units embedded in an elastic matrix and they resist deformation like passive elastic solids; the shape-preserving property makes active solids a superior component for self-driven devices. Owing to the non-equilibrium nature, active solids are predicted to have novel mechanical properties not permitted in passive solids [20-22]. However, active solids have only been investigated in macroscopic analogue models using robotic structures [29-34], and an experimental system appropriate for the study of active solids as continuous media is lacking. Motile bacteria are premier experimental systems for active matter studies [1]. We envisaged that motile bacteria embedded in elastic polymer matrices may constitute a continuum body of active solid because of the superior scalability, so we turned to bacterial biofilms that consist of densely-packed bacterial cells (>$10^{11}$ cells/mL) encased by cell-derived extracellular polymers [16,17]. After testing several commonly used bacterial species, we identified early-stage *Proteus mirabilis* biofilms [35] as a prospective active solid. Cells extracted from such early-stage *P. mirabilis* biofilms (~0.8 μm in width and ~2 μm in length) typically retained motility instead of having transitioned into a sessile state, while there was prominent production of extracellular amyloid fibrils matrix (Extended Data Figure 1; Methods). In the *P. mirabilis* biofilm, cells are self-propelled by rotating flagellar filaments (~20 nm in diameter and ~5-10 μm in length) appended on cell surfaces, and flagellar rotation is fueled by protonmotive force [36]. The bulk storage modulus (i.e., a measure of the elasticity of cell-matrix assembly) of the early-stage *P. mirabilis* biofilms is higher than the loss modulus and ranges from ~$10^2 - 10^3$ Pa (Extended Data Figure 2; Methods), thus the early-stage *P. mirabilis* biofilms represent a viscoelastic solid [37] soft enough to be compliant with the active stress generated by bacteria (~1 Pa for a cell with single flagellum that generates a propulsive force of ~0.57 pN [38] and acts on an area equivalent to the cross-sectional area of the cell body).

We fabricated early-stage *P. mirabilis* biofilms and measured the motion of mass elements in the system (Methods). The biofilms are quasi-2D and disk-shaped, with the top surface exposed to air and the bottom surface in contact with agar that provides nutrient and substrate adhesion. The system is laterally confined by a rim of immotile cells that express little extracellular polymer matrix (Methods; Extended Data Figure 1a,b). A small fraction of cells embedded in the extracellular polymer matrix of the biofilm were labeled by genetically encoded fluorescent protein and were used as tracers of local mass elements (Methods). Strikingly, we found that mass elements in the biofilms are self-driven into local oscillatory motion with nearly homogeneous frequency across space (Fig. 1a,b; Extended Data Figure 3). In circular disk-shaped biofilms of ~1 mm in diameter (Extended Data Figure 1a; representing quasi-two-dimensional isotropic lateral confinement), the local oscillation of mass elements self-organizes into a pair of topologically-distinct, self-driven global motion modes: In one mode, all mass elements being tracked in the biofilm followed a periodic quasi-circular trajectory in a synchronized manner (Fig. 1a; Video S1, Video S2; Methods), hence the entire system underwent periodic translational motion and we referred to this mode as oscillatory translation; in the other mode, the mass elements followed periodic, synchronized quasi-linear trajectories that can be approximated as concentric circular arcs around the center of the disk (Fig. 1b; Video S3, Video S4), hence the entire system underwent global rotation with periodically switching chirality and this mode is referred to as oscillatory rotation. These results confirmed that the mass elements in the biofilms were confined to orbiting about a fixed equilibrium position, as expected in elastic solids; thus we call the early-stage *P. mirabilis* biofilms as bacterial active solid. While *P. mirabilis* biofilm is our choice of study, we note that the solid-like global motion can also be found in *Serratia marcescens* (both oscillatory translation and rotation modes) and in *Escherichia coli* (oscillatory rotation mode), suggesting the generality of the findings (see Methods for details). Interestingly, active-solid like behavior similar to the oscillatory translation mode reported here was found in macroscopic robotic structures [33], whose boundary conditions may prevent the development of global rotational motion.

The two distinct global motion modes found in the bacterial active solid under isotropic lateral confinement were also evident from the temporal dynamics of spatially averaged collective velocity measured by optical flow technique (Fig. 1c-f; Video S3 and Video S4; Methods). In the oscillatory translation mode, the two orthogonal components of spatially averaged collective velocity decomposed in Cartesian coordinate oscillate periodically with $\pi/2$ (clockwise, CW) or $3\pi/2$ (counterclockwise, CCW) phase shift (Fig. 1E). Statistically, the chirality of oscillatory translational motion appears unbiased (CW, 29 out of 60; CCW, 31 out of 60), indicative of spontaneous chiral symmetry breaking. On the other hand, in the oscillatory rotation mode, the tangential component of spatially averaged collective velocity decomposed in polar coordinate oscillates periodically, while the radial component is negligible (Fig. 1f). Interestingly, we found that the two emergent global motion modes in bacterial active solid had a conspicuous difference in oscillation frequency (Methods), with the frequency in oscillatory translation mode (0.11±0.04 Hz; mean±S.D., N = 60) being ~2-fold of that in the oscillatory rotation mode (0.06±0.02 Hz; mean±S.D., N = 41). This finding will be further discussed below.

Under anisotropic lateral spatial confinement with a major axis of symmetry (such as oval-shaped geometry; Extended Data Figure 1d; Methods), we discovered that the local oscillation of mass elements in bacterial active solids is organized in space with a unique pattern of phase distribution. The collective velocity component perpendicular to the major axis (denoted as the transverse component or $v_y$) resembles a standing wave: The phase distribution in space is discretized into regularly separated domains, with each domain having similar phases (Fig. 2a,b; Video S5; Methods); meanwhile, the instantaneous magnitude of $v_y$ varies in space as a sinusoidal function (Fig. 2c). These features are also evident in the spatiotemporal autocorrelation of $v_y$ along the major axis, which displays a characteristic pattern of a standing wave consisting of periodic, segmented domains with high correlation (Fig. 2d). By contrast, the phase of collective velocity component parallel to the major axis (denoted as the parallel component or $v_x$) does not vary in space (Fig. 2a,b); the spatiotemporal autocorrelation of $v_x$ along the major axis displays a pattern of periodic, horizontal lanes with high correlation (Fig. 2e). Taken together, the local oscillation of mass elements in such bacterial active solids under anisotropic lateral confinement is organized in space as a self-sustained transverse standing wave. We note that the phase of $v_x$ always differs from that of $v_y$ by π/2 or –π/2, thus every mass element undergoes oscillatory elliptical motion.

To further understand the findings, we controlled the activity of mass elements in bacterial active solid by tuning the speed of cells with violet light illumination [39] (Extended Data Figure 4; Video S8; Methods). For bacterial active solids under disk-like isotropic lateral confinement, we found that the oscillatory translation and rotation mode dominates at higher and lower activity, respectively. Surprisingly, the modes transit to each other abruptly at certain activity threshold (Fig. 3a,b) with a sharp, step-like frequency jump (a fold change of 2.04±0.42; mean±SD, N = 8) at mode transition (Fig. 3c), which agrees with the ~2-fold frequency difference of the two global motion modes that naturally emerged in disk-shaped bacterial active solids. These results show that activity selectively excites the two global motion modes, revealing a unique emergent mechanical property of active solids. Interestingly, the oscillation frequency of both modes is positively correlated with activity (Fig. 3c). Such activity-dependence of oscillation frequency is also evident in the self-sustained standing waves in bacterial active solids under anisotropic lateral spatial confinement, where the frequency of standing waves increases with activity in a continuous manner (Fig. 3d) rather than taking discrete values as in passive elastic plates. The activity-dependence of oscillation frequency is another unusual mechanical property of active solids.

To rationalize our experimental results, we performed numerical modeling of active solids. We modified the boundary conditions of a particle-based model to describe active solids [18,40,41] by considering a collective of overdamped self-propelled particles connected by Hookean springs (with spring constant $k_b$) and initially arranged in a two-dimensional triangular lattice (Extended Data Figure 5a). Each particle represents a mass element consisting of ~1000 cells (Methods) and has an intrinsic self-propulsion polarity, which accommodates the emergent self-propelled motion of the mass element driven by the propulsive forces of motile cells embedded in the mass element (Methods). Due to substrate adhesion, the particles experience an elastic restoring force pointing towards their initial equilibrium positions, and particles initially sitting at the edge experience an additional elastic force pointing radially towards the edge to account for the steric effect of the lateral confinement boundary (Methods; Extended Data Figure 5a). The position $\vec{x}_i$ and the self-propulsion polarity $\vec{n}_i$ of the $i$-th particle evolve according to the following governing equations (Methods):

$$\dot{\vec{x}}_i = v_0 \vec{n}_i + \Xi_i(\vec{F}_i + D_r \hat{\xi}_r), \qquad (1)$$

$$\dot{\vec{n}}_i = \beta\big[(\vec{F}_i + D_r \hat{\xi}_r) \cdot \hat{n}_i^\perp\big]\hat{n}_i^\perp + D_\theta \hat{\xi}_\theta - \Gamma \frac{\delta F_n}{\delta \vec{n}_i}, \qquad (2)$$

where $v_0$ is the activity of particles and the direction of polarity $\vec{n}_i$ coincides with the direction of particles' self-propelled motion; $\Xi_i$ is a translational mobility tensor; $\vec{F}_i$ is the total external elastic force acting on the particle controlled primarily by the local elasticity of the system (corresponding to spring constants as described in the caption of Extended Data Figure 5A); $\hat{\xi}_r$ and $\hat{\xi}_\theta$ are randomly oriented unit vectors; $\hat{n}_i^\perp$ is a unit vector orthogonal to $\vec{n}_i$; $D_r$, $D_\theta$, $\beta$ and $\Gamma$ are constants. The polarity $\vec{n}_i$ dynamics is controlled by three terms in Eq. (2), including a force-induced reorientation [18], a noise term, and a term involving a Landau-type free energy $F_n = A(-2\vec{n}_i \cdot \vec{n}_i + (\vec{n}_i \cdot \vec{n}_i)^2 +$ $\frac{1}{2}\kappa(\nabla \vec{n}_i)^2)$ [42] that penalizes the deviation of $\vec{n}_i$ from being a unit vector (Methods). The gradient part $\frac{1}{2}\kappa(\nabla \vec{n}_i)^2$ in $F_n$ allows for extending our model to active solids with microscopic geometrical anisotropy and orientational elasticity [20]; nonetheless, here we focused on active solids with isotropic elasticity by setting $\kappa = 0$ in $F_n$. Details of the model are described in Methods.

We found that all active particles in the modeled active solid are self-driven into local oscillatory motion with homogeneous frequency across space. Under two-dimensional isotropic lateral confinement, our simulations successfully reproduced the two topologically distinct global motion modes observed in circular disk-shaped bacterial active solids (Extended Data Figure 5; Video S6, Video S7). We scanned over the parameter space and obtained a phase map for the two global motion modes (Fig. 4a). We found that the emergence of the two global motion modes can not only be selected by particle activity as demonstrated in the experiment, but also by the system's local elasticity; moreover, there was a sharp, step-like frequency jump (~2-fold change) at the phase boundary (Fig. 4b,c; Extended Data Figure 6a), in agreement with the experimental result during activity-controlled mode transition shown in Fig. 3c. Informed by the simulation results, we examined the elasticity dependence of the global motion modes in experiment by controlling the temperature of bacterial active solids that scales linearly with the elasticity of polymer networks [43]. Indeed, we observed a transition between the two global motion modes at certain temperature threshold accompanied by a sharp, ~2-fold frequency change (frequency ratio 2.04±0.77; mean±SD, N = 4) (Extended Data Figure 6B-D). A simulation-informed theoretical analysis provides quantitative insights into the frequency relations uncovered in the two emergent modes; see Supplementary Text for details. Essentially, denoting the frequency in the oscillatory translation and rotation mode as $f_t$ and $f_r$, respectively, the theory yields $f_{t,r}$ being positively correlated with activity ($f_{t,r} \sim \sqrt{v_0}$) and the ratio $f_t/f_t$ independent of activity $v_0$, which are in qualitative agreement with experimental and simulation results.

Under generic anisotropic lateral spatial confinement with a major axis of symmetry (such as in elliptical or rectangular geometry; Extended Data Figure 7a), our simulations also successfully reproduced the self-sustained transverse standing waves found in experiments with bacterial active solids under anisotropic lateral confinement (Fig. 4d;

Extended Data Figure 7; Video S8). The remarkable agreement between experimental and simulation results prompted us to use the active solid model to explore the wave phenomena of active solids in unconfined 2D space, which is currently not attainable in experiments (Methods). We discover that individual particles followed oscillatory quasi-circular motion and the phase of the local oscillation is arranged as traveling waves (Fig. 4e; Video S9). While the frequency $f$ of this traveling wave scales with particle activity $v_0$ as $f \sim v_0^{0.48}$ (Fig. 4f), the wavelength is independent of particle activity (Fig. 4f); thus the wave speed $U$ is proportional to frequency and also scales with activity as $U \sim v_0^{0.48}$. This result is in stark contrast to the counterpart in passive mechanical waves where the wave speed does not depend on the driving amplitude of external stimuli and the wavelength is inversely proportional to frequency in the continuum limit. Notably, the wave speed of such self-generated elastic active matter waves is on the order of a few hundred μ/s estimated based on values measured in the standing wave of bacterial active solids (frequency ~0.1 Hz, wavelength ~1-2 mm), which is much smaller than the speed of acoustic waves in ordinary passive solids. Unlike mechanical waves in passive solids, the elastic waves we uncovered in bacterial active solids are self-generated by the activity of mass elements. Our results illustrate that the inherent active nature of mass elements can give rise to unique properties of active matter waves.

To conclude, we have discovered an array of novel mechanical behaviors of elastic active matter that are not permitted in passive solids, including the formation of self-sustained elastic waves with activity-dependent wave properties and the emergence of two topologically distinct global motion modes with activity-dependent mode selection. The activity-dependence of these uncovered mechanical behaviors provides simple means to control the emergent mechanical properties of solid-state active matter. In addition, our experimental system, i.e., bacterial active solids derived from biofilms, has superior scalability and the convenience of engineering molecule-level interactions using tools from synthetic biology [27,28]; we expect that bacterial active solids will serve as a valuable experimental platform to explore the non-equilibrium physics of active solids in continuum [20-22]. Expanding the framework of classical continuum mechanics [20-22], lessons learned from bacterial active solids may inspire the design of novel functionalities in adaptive or living materials for soft robotics and biotherapeutics [23,24,26-28].

**Supplementary Materials**, including Supplementary Information with details of the theory and Supplementary Videos, is available in the online version of the paper.

**Data availability.** The data supporting the findings of this study are included within the paper and its Supplementary Materials.

**Code availability.** The custom codes used in this study are available from the corresponding author upon request.

**Acknowledgements**. We thank Ye Li and Wenlong Zuo for building the image acquisition and microscope stage temperature control systems, Karine Gibbs (Harvard University) for providing the bacterial strains, Song Liu for helpful discussions, Shiqi Liu for assistance in experiments, and Lei Xu (CUHK) for assistance with bulk rheology measurement. Y.H. and R.Z. thank Qingsong Wang and Jingxuan Tian for fruitful discussions. This work was supported by the National Natural Science Foundation of China (NSFC No. 31971182, to Y.W.), the Research Grants Council of Hong Kong SAR (RGC Ref. No. RFS2021-4S04, 14306820, 14306820; to Y.W.). R.Z. acknowledges financial support from Research Grants Council of Hong Kong SAR (Ref. No. 16300221).

**Author Contributions**: H.X. discovered the phenomena, designed the study, performed experiments, performed simulations, analyzed and interpreted the data. Y.H. and R.Z. developed the theories and improved the simulations. Y.W. conceived the project, designed the study, analyzed and interpreted the data. R.Z. and Y.W. supervised the study. Y.W. wrote the first draft and all authors contributed to the revision of the manuscript.

**Author Information**: Reprints and permissions information is available at www.nature.com/reprints. The authors declare no competing financial interests. Requests for materials should be addressed to Y.W. (ylwu@cuhk.edu.hk).

**Figures**

Figure 1

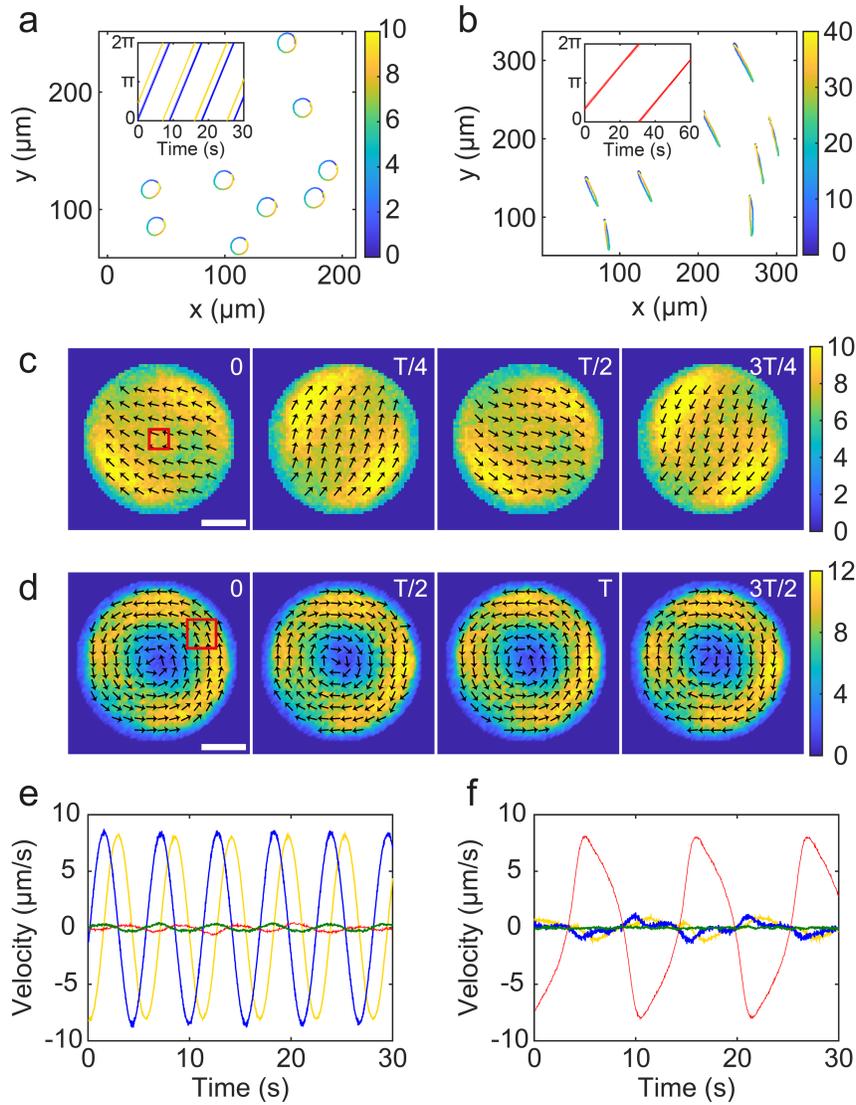

**Fig. 1. Emergent global motion modes in bacterial active solids derived from early-stage *P. mirabilis* biofilms under isotropic lateral confinement. (a,b)** Representative trajectories of mass elements in circular disk-shaped *P. mirabilis* biofilms that underwent global oscillatory translation (panel a) and oscillatory rotation (panel b). The trajectories in each panel were obtained by tracking embedded fluorescent cells in a representative experiment (> 5 replicates; Methods). In panel b, the center of rotation is located to the lower-left direction of the tracking domain. Colormap indicates time (unit: s). Inset of panel a: The position of a mass element was decomposed in Cartesian coordinate as $(x, y)$ and the oscillation phases of the two components (with ~$\pi/2$ phase

difference; Methods) were plotted against time (yellow: $x$; blue: $y$); the phase plots of all mass elements being tracked (N ≥ 20) were shown in an overlaid manner (the plots showed no dispersion because the phases were highly synchronized). Inset of panel b: The position of mass elements was decomposed in polar coordinate and the oscillation phases of the tangential components (Methods) were plotted against time in an overlaid manner (N ≥ 20). (**c,d**) Time sequences of collective velocity field in the global oscillatory translation mode (panel c; period T=5.6 s) and oscillatory rotation mode (panel d; T=11.0 s). The collective velocity field was measured by optical flow analysis based on phase-contrast images (Methods). Arrows represent velocity direction and colormap indicates velocity magnitude. Unit of velocity: $\mu m/s$. Scale bar, 500 $\mu m$. Also see Video S3 and Video S4. Square box at the T= 0 s frame of panel c and d indicates the location of the field of panel a and b, respectively. (**e,f**) Temporal dynamics of spatially averaged collective velocity (Methods) in the global oscillatory translation mode (panel e) and oscillatory rotation mode (panel f). The spatially averaged collective velocity was decomposed as Cartesian (yellow and blue traces) and polar-coordinate components (red: tangential or azimuthal component; green: radial component). In the oscillatory translation mode, the polar-coordinate components are negligible; in the oscillatory rotation mode, both the radial and the Cartesian components are negligible. Panels c,e and d,f are based on data from the same experiments.

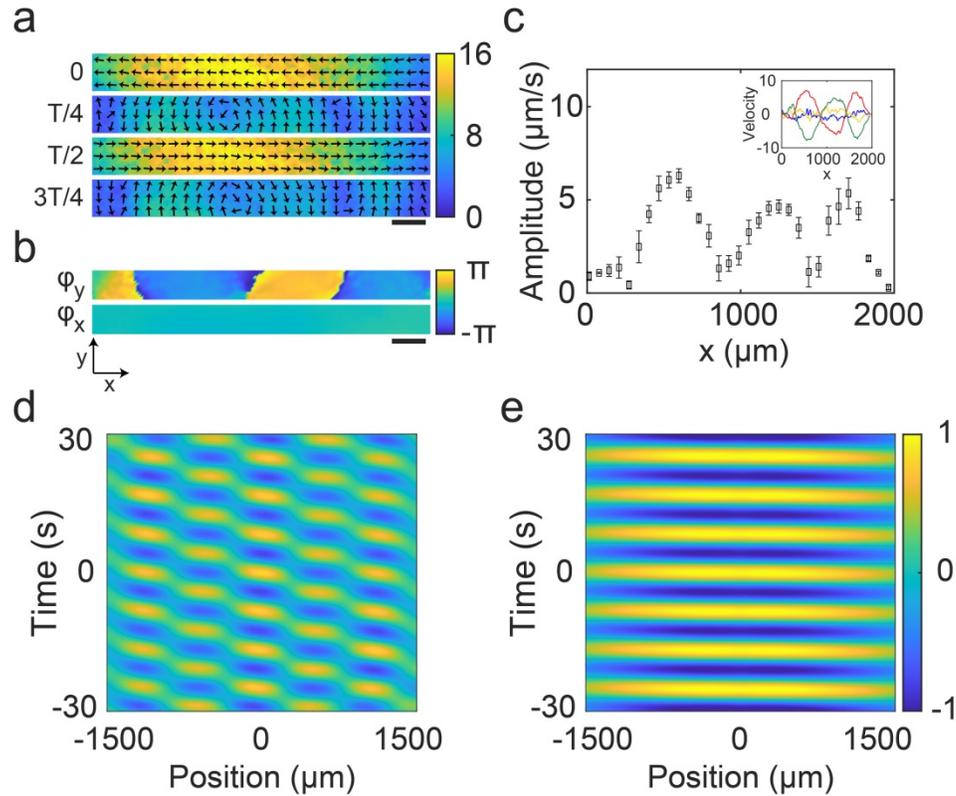

**Fig. 2. Self-sustained transverse standing waves in bacterial active solids under anisotropic lateral confinement.** (**a**) Time sequence of collective velocity field in an oval-shaped bacterial active solid (Extended Data Fig. 1d) that displays the transverse standing wave (with period T=8.6 s) (Methods). The longer side of the rectangular domain shown here is in parallel to the major axis of the bacterial active solid. Arrows represent velocity direction and colormap indicates velocity magnitude. Unit of velocity: μm/s. Also see Video S5. (**b**) Spatial distribution of the oscillation phase of orthogonal collective velocity components associated with panel a. The phase of parallel ($v_x$) and transverse ($v_y$) collective velocity component is denoted as $\varphi_x$ (lower) and $\varphi_y$ (upper), respectively. The positive $x$-axis in the specified coordinate system is parallel to the major axis of the active solid. Scale bars in panels a,b, 200 μm. (**c**) Amplitude distribution of $v_y$ along the major axis of the bacterial active solid. Error bars indicate standard deviation of the velocity amplitude averaged over time (>60 s). Inset: temporal evolution of $v_y$ profile along the major axis, with colors representing the time associated with panel a (blue: 0; red: T/4; yellow: T/2; green: 3T/4). (**d,e**) Spatiotemporal

autocorrelation of $v_y$ (panel d) and $v_x$ (panel e) along the major axis of the bacterial active solid (Methods). Colormap to the right of panel e indicates the autocorrelation magnitude (a.u.). Data in this figure is from a representative experiment (>10 replicates).

Figure 3

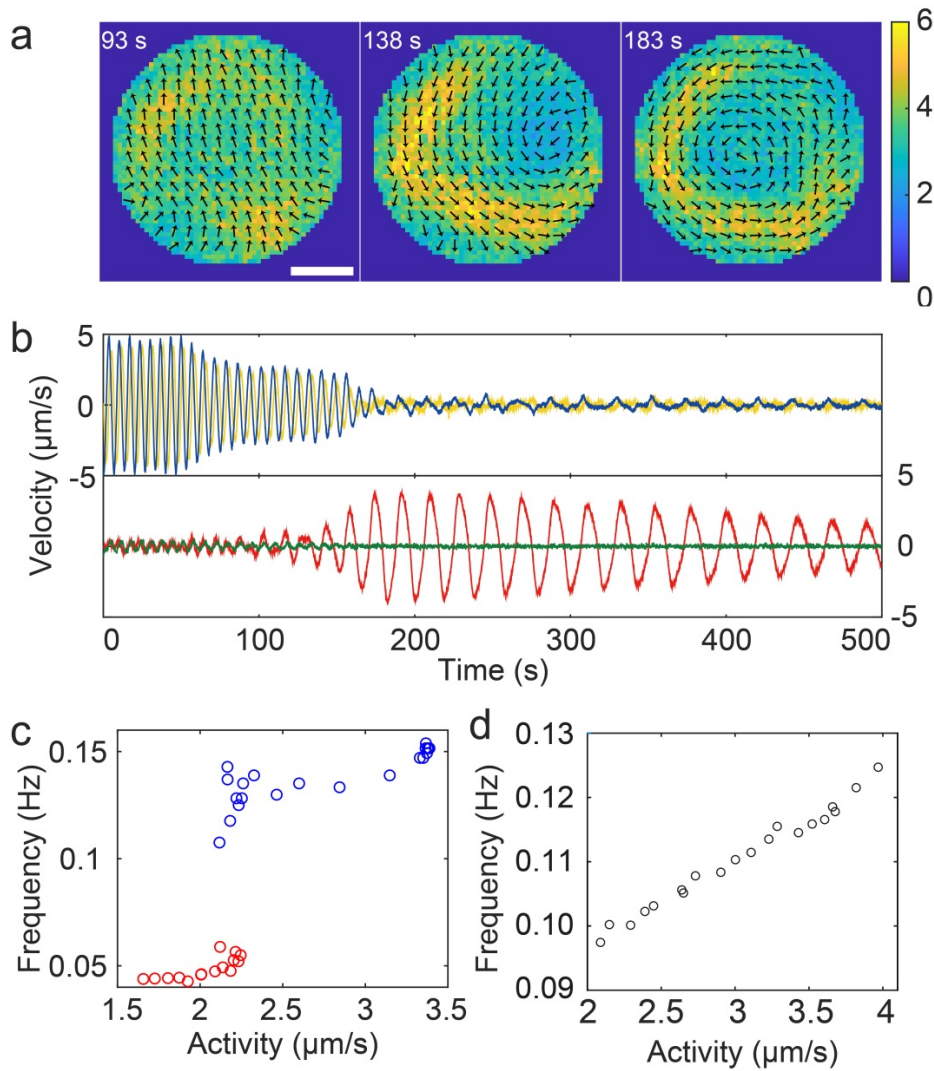

**Fig. 3. Activity selectively excites the global motion modes and controls frequency in bacterial active solids**. (**a,b**) Transition of global motion modes controlled by bacterial motility. A circular disk-shaped bacterial active solid was continuously illuminated by 406 nm violet light starting from T ≈ 60 s (Methods). As cell speed was decreased by violet light (Extended Data Figure 4; Methods), the bacterial active solid underwent mode transition. Panel **a**: Snapshots of collective velocity field in the bacterial active solid during the transition from global oscillatory translation (e.g.,T = 93 s) to global oscillatory rotation (e.g. T = 183 s). The collective velocity field was measured and plotted in the same manner as in Fig. 1c,d. Scale bar, 500 $\mu m$. Panel **b**: Temporal dynamics of spatially averaged collective velocity of the bacterial active solid (Methods) during transition from the oscillatory translation mode to the oscillatory rotation mode

following the decrease of bacterial motility. The spatially averaged collective velocity was decomposed as Cartesian components (yellow and blue traces; upper part of the panel) and polar-coordinate components (red: tangential or azimuthal component, green: radial component; lower part of the panel). (**c**) Activity-dependence of oscillation frequency in the bacterial active solid during mode transition. Horizontal axis represents average collective speed of the active solid (Methods). Color of data points indicates the mode of global motion (blue: oscillatory translation; red: oscillatory rotation). Data in panels a-c were from a representative experiment (>10 replicates). (**d**) Activity-dependence of oscillation frequency in bacterial active solids under anisotropic lateral confinement that display self-sustained transverse standing waves (data from a representative experiment; >10 replicates).

Figure 4

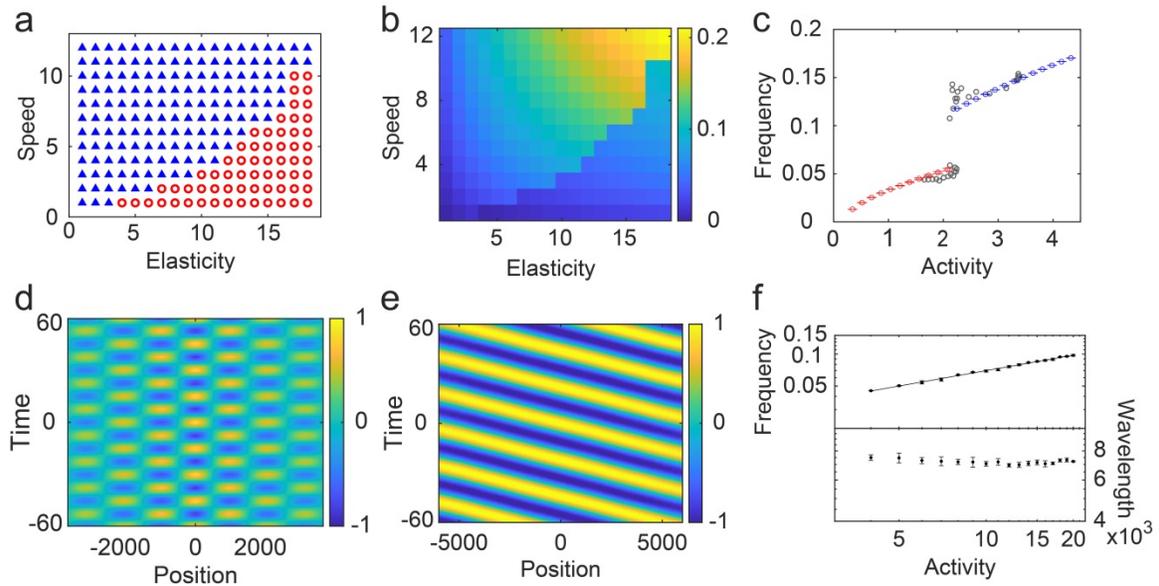

**Fig. 4. Global motion modes and self-sustained elastic waves in modeled active solids.** (**a**) Phase diagram of global motion modes in modeled active solids under isotropic lateral confinement. The phase diagram is plotted in the plane of particles' activity $v_0$ and system's local elasticity (with the inter-particle spring constant $k_b$ serving as a proxy; Methods). Triangles and circles represent oscillatory translation and rotation modes, respectively. Each data point in the phase diagram was obtained with 120 simulation runs (Methods). (**b**) Distribution of oscillation frequency (indicated by the colormap) in the phase diagram of panel a. (**c**) Oscillation frequency of the modeled active solid under isotropic lateral confinement as a function of $v_0$ (fixing $k_b = 14$). Color of data points indicates the mode of global motion (blue: oscillatory translation; red: oscillatory rotation). Error bars indicate standard deviation (N≈100 simulation runs). Empty circles overlaid to the plot are experimental data from Fig. 3c. (**d**) Self-sustained transverse standing waves in a modeled active solid under elliptical lateral confinement. Similar to Fig. 2d, this panel shows the spatiotemporal autocorrelation of the transverse component of particle velocity along the major axis of the elliptical confinement (i.e., the abscissa); the pattern of periodic, segmented domains with high correlation is characteristic of a standing wave. Colormap indicates the autocorrelation magnitude (a.u.). Also see Extended Data Figure 7. (**e**) Self-sustained traveling wave in a modeled active solid in unconfined 2D space. This panel shows the spatiotemporal

autocorrelation of the transverse component of particle velocity along the wave propagation direction (i.e., the abscissa); the pattern of periodic and tilted lanes with high correlation is characteristic of a traveling wave. The component of particle velocity parallel to the wave propagation direction displays the same autocorrelation pattern. Colormap indicates the autocorrelation magnitude (a.u.). (**f**) Oscillation frequency and wavelength in unconfined 2D active solid as a function of particle activity $v_0$. Red line in the upper panel is a linear fit to the log-log plot with a slope of 0.48±0.01 ($R^2$=0.997). Error bars indicate standard deviation (N ≈ 100 simulations runs).

**Extended Data Figures**

Extended Data Figure 1

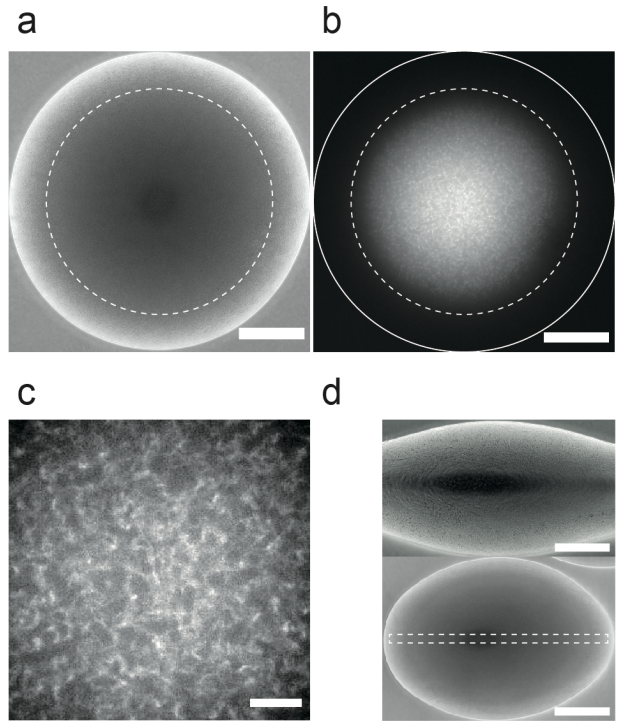

**Extended Data Figure 1. *Proteus mirabilis* colonies at the early stage of biofilm development with prominent production of extracellular polymer matrix**. (**a**) Phase-contrast image of a circular disk-shaped *P. mirabilis* colony grown for 14 hr at 37 °C after inoculation with overnight culture (Methods). Scale bar, 500 $\mu m$. (**b**) Fluorescent image of extracellular amyloid fibrils matrix labeled by Thioflavin T (Methods) in the *P. mirabilis* biofilm shown in panel a. The outer rim of *P. mirabilis* colonies at this development stage is mostly occupied by immotile cells that have transitioned to the sessile state but expressed little extracellular matrix. The width of this outer immotile rim varies from tens to hundreds of µm across different colonies; in the case of panel a, the immotile rim ranges from radius R= ~860 µm (measured from the colony center) to R= 1154 µm (i.e, the colony edge), spanning a width of ~300 µm. The inner region of the colonies with Thioflavin T fluorescence (i.e., the biofilm region; enclosed by the dashed circle in panels a,b) is where we choose to study and refer to as early-stage biofilm or bacterial active solid, such as the fields in main text Fig. 1c,d and 3a. The immotile outer rim of the colony serves as the lateral spatial confinement for the bacterial active solid. Scale bar, 500 $\mu m$. (**c**) Enlarged view of the center of panel b. Scale bar, 100 $\mu m$. (**d**)

Phase-contrast images of oval-shaped *P. mirabilis* colonies with various values of the eccentricity. Scale bars, $500 \mu m$. The field in main text Fig. 2a corresponds to the stripe region enclosed by the dashed rectangle in panel d.

Extended Data Figure 2

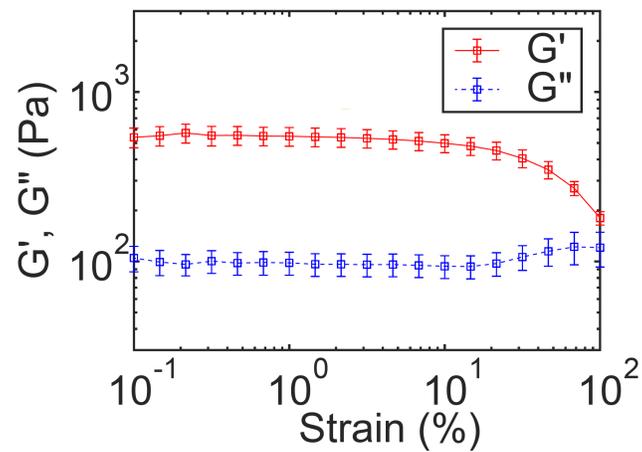

**Extended Data Figure 2. Dynamic moduli of the bacterial active solid derived from early-stage *P. mirabilis* biofilms.** Storage (G'; red) and loss (G''; blue) modulus of the bacterial active solid as a function of strain were measured by bulk rheometry with a cone-plate rheometer in shear-strain-amplitude sweep mode (at constant frequency) (Methods). The oscillation frequency and the temperature of the rheometer were set at 1Hz and 37 °C, respectively. Error bars indicate standard deviation (N=3). The solid and dashed lines are guides to the eye.

Extended Data Figure 3

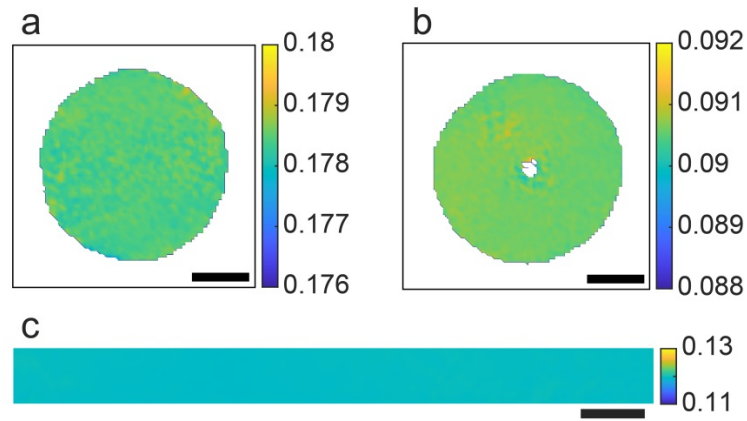

**Extended Data Figure 3.** Spatial distribution of local oscillation frequency in bacterial active solids. Panels a and b shows the frequency distribution in circular disk-shaped bacterial active solids undergoing global oscillatory translation and oscillatory rotation, respectively, with the color bars indicating the magnitude of local oscillation frequency (unit: Hz). Scale bars, 500 $\mu m$. The frequency at the center of panel b is absent because the velocity there is vanishing. Panel c shows the frequency distribution in a rectangular region of an oval-shaped bacterial active solid that displays the transverse standing wave, with the color bar indicating the magnitude of local oscillation frequency (unit: Hz). The longer side of the selected region is parallel to the major axis of symmetry of the bacterial active solid. Scale bar, 200 $\mu m$.

Extended Data Figure 4

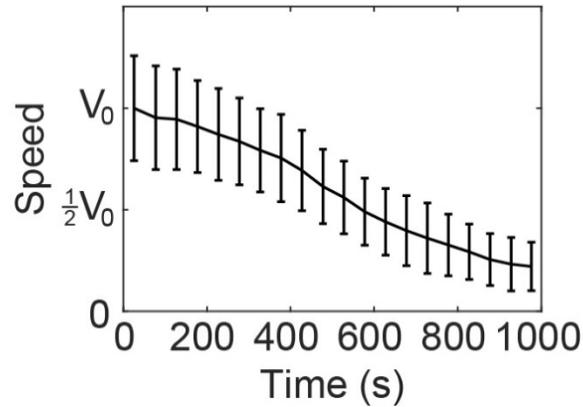

**Extended Data Figure 4. Tuning single-cell speed of *P. mirabilis* via violet light illumination.** To obtain this plot, cells were extracted from the *P. mirabilis*-based bacterial active solids that were undergoing either global oscillatory translation or global oscillatory rotation; the extracted cells were mixed with 0.02% Tween 20 and deposited on 0.6% LB agar surface, forming a quasi-2D dilute bacterial suspension drop. *P. mirabilis* cells in the prepared quasi-2D dilute bacterial suspension drop were continuously illuminated by 406 nm violet light starting from T = 0 s while being tracked in phase-contrast microscopy (Methods); note that for single-cell tracking, a 20x objective lens was used. The speed of an individual cell at a specific time T was computed based on its trajectory tracked from (T-0.5) s to (T+0.5) s; the single-cell speeds computed from (T-25) s to (T+25) s were then averaged and taken to be the mean cell speed at T. Data shown in the plot was normalized by the mean speed at T= 0 s (i.e., the free-swimming speed of cells without blue light illumination; $V_0 = $ 26.7±6.9 μm/s; mean±S.D., N ≈ 2500). Error bars indicate standard deviation (N ≈ 2500).

Extended Data Figure 5

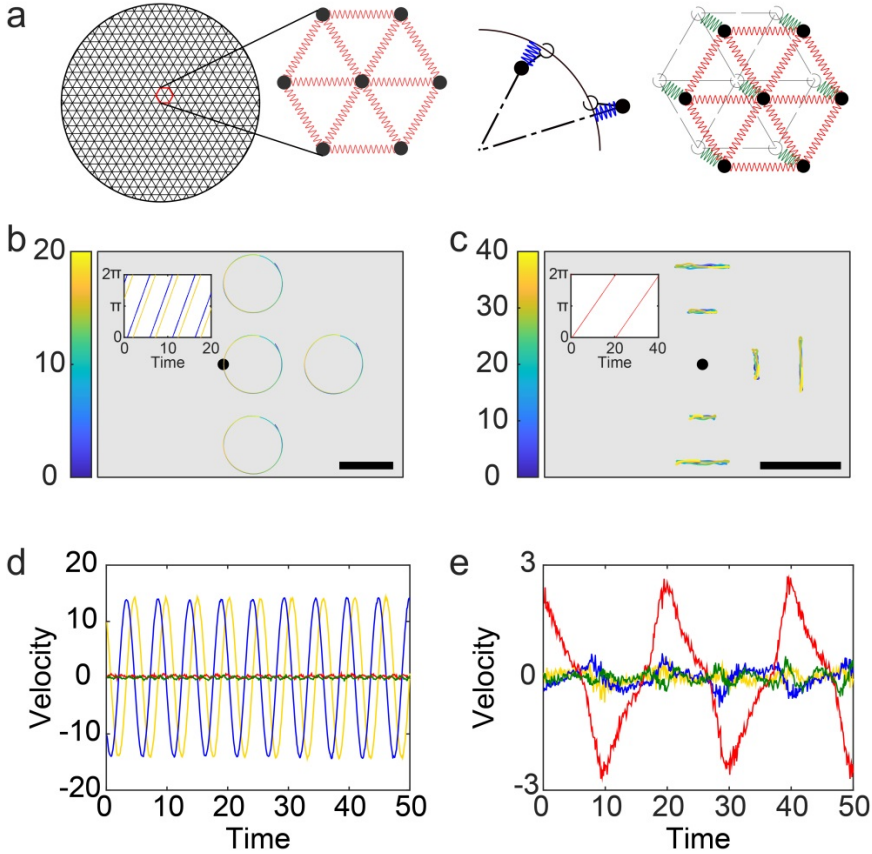

**Extended Data Figure 5. Emergent global motion modes in modeled active solid under isotropic lateral confinement.** **(a)** Schematic diagram of the bead-spring model for active solid under two-dimensional isotropic lateral confinement (Methods). The model consists of $N = 511$ self-propelled particles (black solid circles). Every nearest-neighbor pair of particles is connected by an inter-particle spring with spring constant $k_b$ (red). The particles also experience elastic forces due to substrate adhesion and lateral spatial confinement (see main text) via a restoring spring (green) and a boundary spring (blue) with spring constant $k_s$ and $k_r$, respectively. The three spring constants together determine the system's local elasticity. In simulations the inter-particle spring constant $k_b$ was used as a proxy for the system's local elasticity, with the ratios between $k_b$, $k_s$ and $k_r$ fixed. **(b,c)** Representative trajectories of particles in the modeled active solid that underwent global oscillatory translation (panel b) and oscillatory rotation (panel c). Most particles (except those very near the center or the boundary) followed periodically oscillating quasi-circular trajectories (at relatively high activity; panel b; Video S6) or

quasi-linear concentric trajectories (at relatively low activity; panel c; Video S7) with highly synchronized phases (insets of panel b,c), in the same manner as the motion of matrix-embedded cells in the experiments undergoing global oscillatory translation (main text Fig. 1a) or rotation (main text Fig. 1b), respectively. Black dot in each panel indicates the center of the simulation domain. Scale bars represent 1/3 of the inter-particle distance at equilibrium and colormap indicates time. Insets: Oscillation phases of individual particle's velocity components plotted in the same way as in insets of Fig. 1a,b. Simulation parameters: $v_0 = 15$ (panel b) or $v_0 = 3$ (panel c), $k_b = 12, k_r = 0.6$, and $k_s = 0.044$. (**d,e**) Temporal dynamics of spatially averaged particle velocity in the modeled active solid in global oscillatory translation mode (panel d) or oscillatory rotation mode (panel e). The velocity was averaged over all particles in the simulation and then decomposed as Cartesian (yellow and blue traces) and polar-coordinate components (red: tangential or azimuthal component; green: radial component). The spatially averaged particle velocity in the two emergent modes was characterized by distinct temporal dynamics in Cartesian or polar coordinates similar to that found in the experiments (main text Fig. 1e,f). Simulation parameters are identical to those used in panels b,c.

Extended Data Figure 6

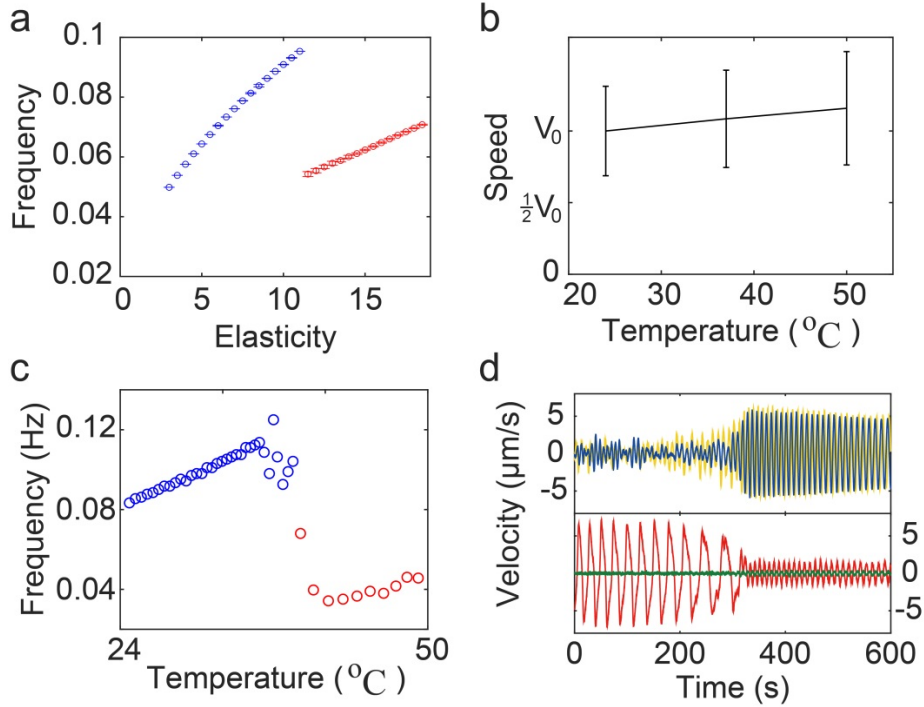

**Extended Data Figure 6. Elasticity dependence of the global motion modes in active solids under isotropic lateral confinement.** (**a**) Oscillation frequency of global motion modes in modeled active solids as a function of $k_b$ (fixing $v_0 = 4$). Color of data points indicates the mode of global motion (blue: oscillatory translation; red: oscillatory rotation). Error bars indicate standard deviation (N≈100 simulation runs). (**b**) Temperature dependence of *P. mirabilis* single-cell speed. *P. mirabilis* cells in quasi-2D dilute bacterial suspension drops (prepared in the same manner as described in the caption of Extended Data Figure 4) were tracked in fluorescent microscopy while the environmental temperature was varied from 24 °C to 50 °C with a custom-built temperature-control system (Methods). As shown in the plot the speed of cells only changed slightly in this temperature range (up to ~15%). The mean speed of cells at a specific temperature was computed based on 1-s segments of cell trajectories tracked in a 200-s time window. Data shown in the plot was normalized by the mean speed at temperature 24 °C. Error bars indicate standard deviation (N ≈ 2500). (**c,d**) Transition of global motion modes in bacterial active solids controlled by temperature. Panel c shows the temperature-dependence of oscillation frequency in the bacterial active solid during mode transition. Color of data points indicates the mode of global motion (blue: oscillatory translation; red: oscillatory rotation). Panel d shows the temporal dynamics of

spatially averaged collective velocity during transition from the oscillatory rotation mode to the oscillatory translation mode following the decrease of temperature. The spatially averaged collective velocity was decomposed as Cartesian components (yellow and blue traces; upper part of panel d) and polar-coordinate components (red: tangential or azimuthal component, green: radial component; lower part of panel d). Data in panels c,d were from a representative experiment (>5 replicates).

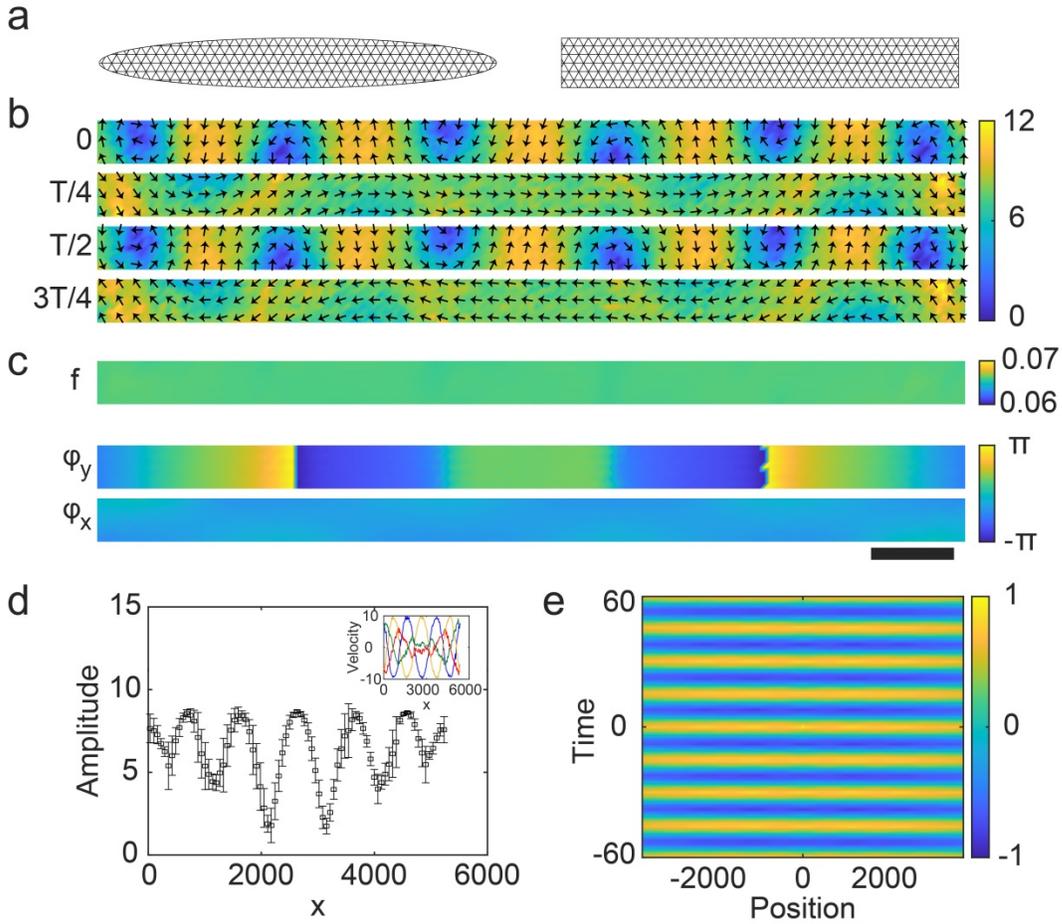

**Extended Data Figure 7. Self-sustained transverse standing waves in modeled active solids under anisotropic lateral confinement.** This figure is associated with main text Fig. 4d. Simulation parameters: $v_0 = 10$, $k_b = 4$, $k_r = 0.2$, and $k_s = 0.015$. (**a**) Schematic diagrams of the bead-spring model for active solid under two-dimensional anisotropic lateral confinement (Methods). The elliptical confinement (left) mimics the anisotropic lateral confinement geometry of oval-shaped bacterial active solids used in main text Fig. 2. Rectangular lateral confinement (right) produces similar simulation results. (**b**) Time sequence of particle velocity field in a modeled active solid under elliptical lateral confinement that displays the transverse standing wave. The longer side of the rectangular domain shown here is in parallel to the major axis of the elliptical confinement. T denotes the period of oscillation. Arrows represent velocity direction and colormap indicates velocity magnitude. (**c**) Spatial distributions of local oscillation frequency and phase associated with panel b. The local oscillation frequency is

homogeneous in space (upper part of the panel, with the color bar indicating the magnitude of frequency). The phase of parallel ($v_x$; parallel to the major axis) and transverse ($v_y$; perpendicular to the major axis) component of particle velocity is denoted as $\varphi_x$ (lower) and $\varphi_y$ (upper), respectively. Scale bar under panel c is shared by panel b and represents 7.7 times of the inter-particle distance at equilibrium. (**d**) Averaged amplitude distribution of $v_y$ along major axis of the modeled active solid. Error bars indicate standard deviation of the velocity amplitude averaged over >10 simulation runs. Inset: temporal evolution of $v_y$ profile in panel b along the major axis of the modeled active solid, with colors representing time (blue: 0; red: T/4; yellow: T/2; green: 3T/4). (**e**) Spatiotemporal autocorrelation of $v_x$ along the major axis of the modeled active solid (i.e., the abscissa of the figure); the pattern of periodic, horizontal lanes with high correlation is similar to that seen in experiment (main text Fig. 2e). Colormap at right side indicates the autocorrelation magnitude (a.u.).

Extended Data Figure 8

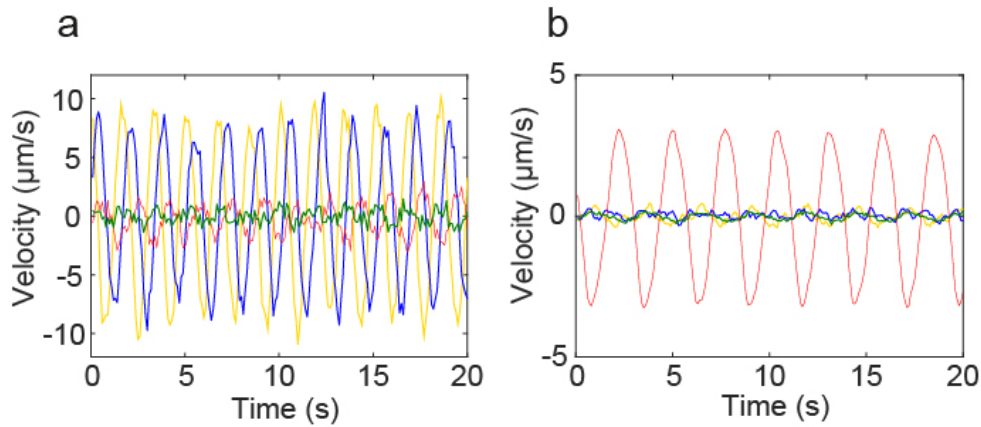

**Extended Data Figure 8. Emergent global motion modes in bacterial active solids derived from *S. marcescens* biofilms.** Temporal dynamics of spatially averaged collective velocity in the global oscillatory translation mode (panel a) and oscillatory rotation mode (panel b). The spatially averaged collective velocity was decomposed as Cartesian (yellow and blue traces) and polar-coordinate components (red: tangential or azimuthal component; green: radial component). In the oscillatory translation mode, the polar-coordinate components are negligible; in the oscillatory rotation mode, both the radial and the Cartesian components are negligible.

Extended Data Figure 9

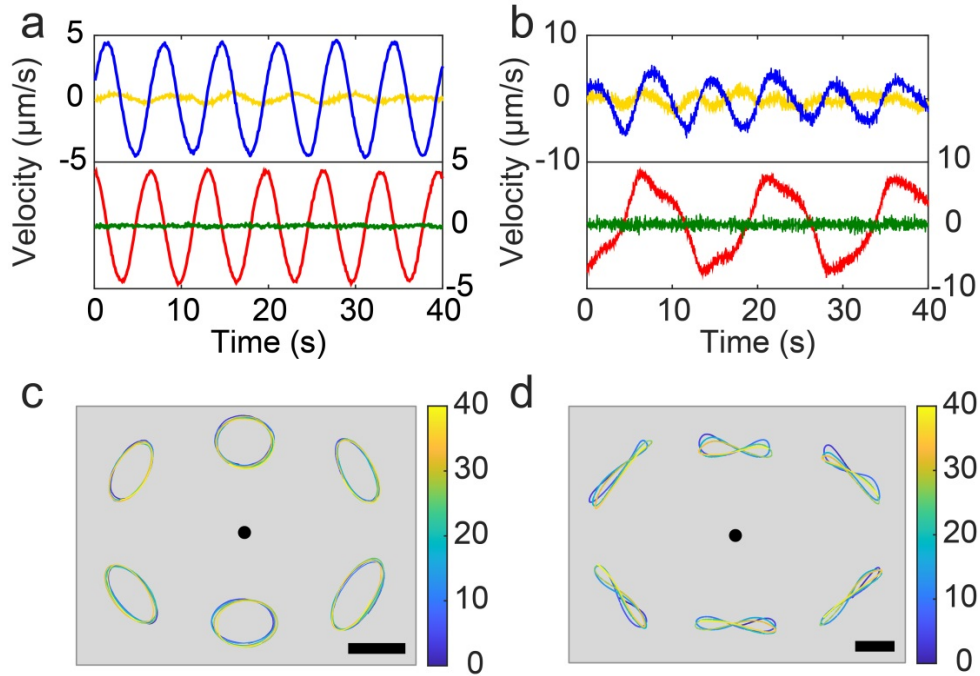

**Extended Data Figure 9. Coexistence of two global motion modes in bacterial active solids under isotropic lateral confinement.** (**a,b**) Temporal dynamics of spatially averaged collective velocity in circular disk-shaped bacterial active solids where the oscillatory translation and oscillatory rotation modes co-existed with identical frequencies ($f_t = f_r = 0.15$ Hz; panel a) or with the frequency of the oscillatory translation mode doubled ($f_t = 0.14$ Hz, $f_r = 0.07$ Hz; panel b). The velocity was decomposed as Cartesian (yellow and blue traces) and polar-coordinate components (red: tangential component; green: radial component). (**c,d**) Panels c and d display representative trajectories of mass elements in bacterial active solids analyzed in panels a and b, respectively. In each panel the trajectories were obtained by integrating the spatially averaged collective velocity over a 50 $\mu m$ × 50 $\mu m$ domain located ~ 500 µm from the center of the disk-shaped bacterial active solid (black dot) at different polar angles. The trajectories were brought close to the center for better visualization, and thus the scale bars (panel c, 10 $\mu m$; panel d, 20 $\mu m$) apply to the trajectories only. Colormap indicates time (unit: s).

Extended Data Figure 10

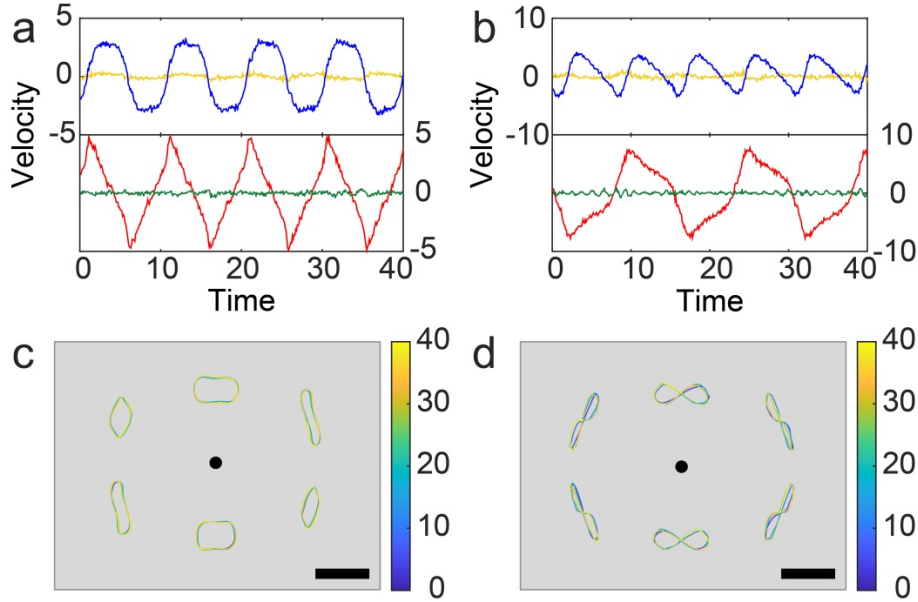

**Extended Data Figure 10. Coexistence of two global motion modes in modeled active solid under isotropic lateral confinement.** (**a,b**) Temporal dynamics of collective velocity of the system where oscillatory translation and oscillatory rotation modes co-existed with identical frequencies ($f_t = f_r = 0.10$ ; panel a) or with the frequency of the oscillatory translation mode doubled ($f_t = 0.13$, $f_r = 0.07$; panel b). The collective velocity was averaged over all particles in the simulation and then decomposed as Cartesian (yellow and blue traces; upper part of each panel) and polar-coordinate components (red: tangential or azimuthal component, green: radial component; lower part of each panel). Simulation parameters: panel a, $v_0 = 5, k_b = 12$; panel b, $v_0 = 9, k_b = 16$; the ratios between $k_b$, $k_s$ and $k_r$ are fixed (Methods). (**c,d**) Panels c and d display the trajectory of representative particles in the simulations analyzed in panels a and b, respectively. In each panel the particle was chosen at ~5 times of the equilibrium inter-particle distance from the center of the circular simulation domain (black dot) at different polar angles. The trajectories were brought close to the center for better visualization, and thus the scale bars (indicating 1/3 inter-particle distance at equilibrium) apply to the trajectories only. Colormap indicates time.